# Bound and scattering state of position dependent mass Klein-Gordon equation with Hulthen plus deformed-type hyperbolic potential

*A N Ikot*[1], *H.P.Obong* [1] ,*T M Abbey* [1], *M Ghafourian* [2] *and H Hassanabadi* [3]

[1] Theoretical Physics Group,Department of Physics, University of Port Harcourt, Choba,P.M.B 5323 Port Harcourt-Nigeria.
2 Department of Basic Sciences, Islamic Azad University, North Tehran Branch, Tehran, Iran
3 Department of Physics,Shadrood University-Iran

**email:ndemikotphysics@gmail.com*
*Tel: +2348038767186*

**Abstract**

In this article we used supersymmetry quantum mechanics and factorization methods to study the bound and scattering state of Klein-Gordon equation with deformed Hulthen plus deformed hyperbolical potential for arbitrary state in D-dimensions. The analytic relativistic bound state eigenvalues and the scattering phase factor are found in closed form. We reported on the numerical results for the bound state energy in D-dimensions.

## 1. Introduction

In recent times there has been a considerable effort by different authors to investigate the analytical solutions of the Klein-Gordon equation within the framework of higher spatial dimensions or D-dimensions with motivated physical potential models [1-5]. The famous Klein-Gordon equation in relativistic quantum mechanics is one of the most frequently used equations that describes spin zero particles such as mesons [6-8]. The exact or approximate solutions of the Klein-Gordon equation are very important in physics and chemistry because its solution contain all the necessary information needed for the complete description of the quantum state of the system such as probability density and entropy [9-12] and this quantum system is exactly solvable if all the eigenvalues and eigenfunctions can be calculated analytically [13-15]. In order to obtain the exact and approximate solutions of the Klein-Gordon equation, various quantum mechanical techniques have been employed such as Nikiforov-Uvarov(NU) [16], Supersymmetric quantum mechanics(SUSYQM)[17], asymptotic interation method(AIM)[18],exact quantization rule[19] and others[20]. Jia et al [21] studied the Klein-Gordon equation with improved Manning-Rosen potential using SUSYQM. Xie and Jia [22] investigate the Klein-Gordon equation in higher dimensions with Morse potential. Ikot et al [23] studied the Klein-Gordon equation in D-dimensions for multiparameter exponential typed potential. Many other researchers have investigated the solutions of the wave equation with different potential models such as Hylleraas potential[24],Mobius square potential[25],Kratzer potential[26],Rosen –Morse potential[27], new generalized Morse potential [28] and pseudoharmonic potential [29]. Recently, Ikot et al. [30] studied the D-dimensional Klein-Gordon equation with improved Manning-Rosen potential. Hassanabadi and Yazarloo[31] studied the bound and scattering state solutions of Klein-Gordon equation with generalized Poschl-Teller potential. The bound and scattering state solutions of Schrödinger equation with second Poschl-Teller potential have been investigated by You et al.[32].The scattering state of the coupled Woods-Saxon potential for Duffin Kemmer Petiau(DKP) equation had been examined by Ikot et al.[33]. Arda et al.[34] have investigated the bound and scattering state solutions of the effective mass Klein Gordon equation. Chen et al.[35] reported on the scattering state of Coulomb potential plus ring shaped potential for Klein-Gordon equation. Different approximation scheme has been employed to study the bound state solutions of the wave equation for arbitrary $l \neq 0$ state[36-37]. Many authors used the approximation schemes [37-40] (see fig.1)

$$\frac{1}{r^2} \approx 4\alpha^2 \left[ c_0 + \frac{e^{-2\alpha r}}{1-e^{-2\alpha r}} + \left( \frac{e^{-2\alpha r}}{1-e^{-2\alpha r}} \right)^2 \right] \qquad (1)$$

for the centrifugal term to obtain the approximate solutions of the bound state of Schrödinger , Klein-Gordon and Dirac equation using various methods of NU,AIM and SUSYQM. As noted by You et al.[32],this approximation for the centrifugal term cannot be used to obtain the scattering state because of the constant $c_0$ but can be used effectively for the study of the bound state solutions. It is well-known that the position dependent mass (PDM) solutions of relativistic and non-relativistic quantum mechanics are of great important in physics and have attracted a considerable attention in recent times[41-42]. Many authors have studied the PDM for Schrödinger, Klein-Gordon and Dirac equations with different potentials [43-45].

The aim of this paper is to obtain the bound and scattering states energy eigenvalues, phase shift and the normalization the eigen function for PDM Klein-Gordon equation with Hulthen plus deformed hyperbolic potential in D-dimension. The Hulthen plus doformed-type hyperbolic potential is defined as [46](see fig.2)

$$V(r) = -V_0 \frac{e^{-2\alpha r}}{1-qe^{-2\alpha r}} + V_1 \coth_q\left(\alpha r\right) \qquad (2)$$

where, $V_0$ and $V_1$ are the depth of potential wells, $q$ is the deformation parameter and $\alpha$ is inverse of screening parameter (see fig.2). The mass function is defined as [47],

$$m(r) = m_0 + \frac{m_1}{1-qe^{-2\alpha r}} \qquad (3)$$

where $m_0$ is the rest mass of relativistic particles and $m_1$ is the perturbed mass. The following can be deduced from the mass function as special cases,

$$m(r) = \begin{cases} m_0, m_1 \to 0 \\ m_0 + m_1, \alpha \to \infty \\ m_0 + \dfrac{m_1}{1-q}, \alpha \to 0 \end{cases} \qquad (4)$$

The $q$-parameter used in the deformed hyperbolic potential are defined as follows [48],

$$\sinh_q(x) = \frac{e^x - e^{-x}}{2}, \cosh_q(x) = \frac{e^x + e^{-x}}{2},$$
$$\tanh_q(x) = \frac{\sinh_q(x)}{\cosh_q(x)}, \csc h_q(x) = \frac{1}{\sinh_q(x)}, \qquad (5)$$
$$\cosh_q^2(x) - \sinh_q^2(x) = q$$

**2. Bound state Solution of Klein-Gordon Equation in D-Dimension**

The Klein-Gordon equation in higher dimension for spherically symmetric potential reads [49-50],

$$-\hbar^2c^2\Delta_D\psi_{n,l,m}(r,\Omega_D)=\left\{[E_{n,l}-V(r)]^2-[mc^2+S(r)]^2\right\}\psi_{n,l.m}(r,\Omega_D) \qquad (6)$$

Where $E_{n,l},m,V(r)$ and $S(r)$ are the relativistic energy ,rest mass, the repulsive vector potential and the attractive scalar potential respectively and $\Delta_D$ is defined as

$$\Delta_D=\nabla_D^2=\frac{1}{r^{D-1}}\frac{\partial}{\partial}\left(r^{D-1}\frac{\partial}{\partial r}\right)-\frac{\Lambda_D^2(\Omega_D)}{r^2} \qquad (7)$$

The total wave function in D-dimension is written as,

$$\psi_{n,l,m}(r,\Omega_D)=R_{n,l_D}(r)Y_{l_D}^m(\Omega_D) \qquad (8)$$

The term $\frac{\Lambda_D^2(\Omega_D)}{r^2}$ is the generalization of the centrifugal term for the higher dimensional space. The eigenvalues of $\Lambda_D^2(\Omega_D)$ are defined by the relation,

$$\Lambda_D^2(\Omega_D)Y_{l_D}^m(\Omega_D)=l_D(l_D+D-2)Y_{l_D}^m(\Omega_D) \qquad (9)$$

Where $Y_l^m(\Omega_D),R_{n,l}$ and $l_D$ represent the hyperspherical harmonics, the hyperradial wave function and the orbital angular momentum quantum number respectively. Now substituting ansatz $R_{n,l_D}(r)=r^{-\frac{(D-1)}{2}}F_{n,l_D}(r)$ for the wave function into equation (2) yields,

$$\left\{\hbar^2c^2\frac{d^2}{dr^2}+\left(E_{n,l}-V(r)\right)^2-\left(m(r)c^2+S(r)\right)^2-\frac{(D+2l_D-1)(D+2l_D-3)\hbar^2c^2}{4r^2}\right\}F_{n,J_D}(r)=0 \qquad (10)$$

Considering unequal scalar and vector potential for the Hulthen plus deformed hyperbolic potential as,

$$V(r)=-V_0\frac{e^{-2\alpha r}}{1-qe^{-2\alpha r}}+V_1\left(\frac{1+qe^{-2\alpha r}}{1-qe^{-2\alpha r}}\right),$$
$$S(r)=-S_0\frac{e^{-2\alpha r}}{1-qe^{-2\alpha r}}+S_1\left(\frac{1+qe^{-2\alpha r}}{1-qe^{-2\alpha r}}\right) \qquad (11)$$

we obtain the following second-order Schrödinger-like equation

$$\left\{\hbar^2c^2\frac{d^2}{dr^2}+\left(E_{n,J_D}-\left(-V_0\frac{e^{-2\alpha r}}{1-qe^{-2\alpha r}}+V_1\left(\frac{1+qe^{-2\alpha r}}{1-qe^{-2\alpha r}}\right)\right)\right)^2\right\}U_{n,l_D}(r)$$
$$-\left[\left(\left(m_0+\frac{m_1}{1-qe^{-2\alpha r}}\right)c^2+\left(-S_0\frac{e^{-2\alpha r}}{1-qe^{-2\alpha r}}+S_1\left(\frac{1+qe^{-2\alpha r}}{1-qe^{-2\alpha r}}\right)\right)\right)^2\right]U_{n,l_D}(r)$$
$$-\left(\frac{(D+2l_D-1)(D+2l_D-3)\hbar^2c^2}{4r^2}\right)U_{n,l_D}(r)=0 \qquad (12)$$

It is well-known that equation (12) cannot be solved exactly except for $J=0$. In order to find the solutions of Eq.(12), we apply a suitable approximation for the centrifugal term[48] since approximation in equation (1) is not suitable for both bound and scattering state problem. We invoke the suitable approximation as[48] (see fig.3)

$$\frac{1}{r^2}\approx\frac{\alpha^2}{\sinh_q^2(\alpha r)}=\frac{4\alpha^2}{\left(1-qe^{-2\alpha r}\right)^2} \qquad (13)$$

Substituting Eq,(13) into Eq,(12) and after a little algebra, we get

$$-\frac{d^2F_{nJ_D}}{dr^2}+\frac{1}{\left(1-qe^{-2\alpha r}\right)^2}\left[\omega_1e^{-4\alpha r}+\omega_2e^{-2\alpha r}+\omega_3\right]F_{n,J_D}(r)=\tilde{E}_{nJ_D}F_{nJ_D} \qquad (14)$$

Where,

$$\tilde{E}_{n,J_D}=\frac{E_{n,J_D}^2-m^2c^4}{\hbar^2c^2} \qquad (15)$$

$$\omega_1=\frac{1}{\hbar^2c^2}\left[2m_0c^2S_0q-2m_0c^2S_1q^2+2E_{n,J_D}V_0q-2E_{n,J_D}V_1q^2+S_0^2+S_1^2q^2-V_0^2+V_1^2q^2\right]$$

$$\omega_2=\frac{1}{\hbar^2c^2}\left[-2m_0c^2S_0-2m_1c^2S_0+2m_1c^2S_1q-2E_{nJ_D}V_0-2S_0S_1q+2S_1^2q+2V_0V_1q+2V_1^2q-2m_0m_1c^4q+4\gamma\hbar^2c^2\alpha^2\right]$$

$$\omega_3=\frac{1}{\hbar^2c^2}\left[2m_1c^2S_1+2m_0c^2S_1+2E_{nJ_D}V_1-2S_0S_1+S_1^2+2V_0V_1+V_1^2+2m_0m_2c^4+m_1^2c^4\right]$$
$$\gamma=\frac{(D+2l_D-1)(D+2l_D-3)}{4} \qquad (16)$$

In the SUSYQM formulation, the ground-state wave function $F_{0,l}(r)$ is given by [17,21,30]

$$F_{0,l}(r)=\exp\left(-\int W(r)dr\right), \quad (17)$$

in which the integrand is called the superpotential and the Hamiltonian is composed of the raising and lowering operators

$$H_- = \hat{A}^+\hat{A} = -\frac{d^2}{dr^2} + V_-(r), \qquad (18)$$

$$H_+ = \hat{A}\hat{A}^+ = -\frac{d^2}{dr^2} + V_+(r), \qquad (19)$$

with

$$\hat{A} = \frac{d}{dr} - W(r), \qquad (20)$$

$$\hat{A}^+ = -\frac{d}{dr} - W(r), \qquad (21)$$

$$V_\pm(r) = W^2(r) \mp W'(r) \qquad (22)$$

Now substituting Eq.(17) into Eq.(14), we obtain the associated Riccati equation:

$$W^2(r) - W'(r) = \frac{1}{\left(1 - qe^{-2\alpha r}\right)^2}\left[\omega_1 e^{-4\alpha r} + \omega_2 e^{-2\alpha r} + \omega_3\right] - \tilde{E}_{0,l}, \qquad (23)$$

Here, we choose the superpotential of the form,

$$W(r) = P + \frac{Qe^{-2\alpha r}}{\left(1 - qe^{-2\alpha r}\right)} \qquad (24)$$

Substituting Eq.(24) into Eq.(23) and solve explicitly, we obtain the following

$$P = -\left[\frac{-Q^2 + \omega_1 - \omega_3 q^2}{2qQ}\right] \qquad (25)$$

$$Q = \alpha q\left[-1 \pm \sqrt{1 + \frac{1}{\alpha^2 q^2}\left\{\omega_1 + \omega_2 q + \omega_3 q^2\right\}}\right] \qquad (26)$$

$$\tilde{E}_{0,J_D} = -\left(\frac{-Q^2 + \omega_1 - \omega_3 q^2}{2qQ}\right)^2 + \omega_3 \qquad (27)$$

We construct the pair of supersymmetric partner potentials $V_+(r)$ and $V_-(r)$ as follows,

$$V_+(r) = W^2(r) + \frac{dW(r)}{dr},$$

$$= \frac{Q\left(Q - 2\alpha\right)e^{-2\alpha r}}{\left(1 - qe^{-2\alpha r}\right)^2} + \frac{Q^2 e^{-4\alpha r} - Q^2 e^{-2\alpha r}}{\left(1 - qe^{-2\alpha r}\right)^2} + \frac{2PQe^{-2\alpha r}}{\left(1 - qe^{-2\alpha r}\right)} + \left(\frac{-Q^2 + \omega_1 - \omega_3 q^2}{2qQ}\right)^2 \qquad (28)$$

$$V_-(r) = W^2(r) - \frac{dW(r)}{dr}, \qquad (28)$$

$$= \frac{Q\left(Q + 2\alpha\right)e^{-2\alpha r}}{\left(1 - qe^{-2\alpha r}\right)^2} + \frac{Q^2 e^{-4\alpha r} - Q^2 e^{-2\alpha r}}{\left(1 - qe^{-2\alpha r}\right)^2} + \frac{2PQe^{-2\alpha r}}{\left(1 - qe^{-2\alpha r}\right)} + \left(\frac{-Q^2 + \omega_1 - \omega_3 q^2}{2qQ}\right)$$

The partner potential are shape invariance via mapping of the form $Q \to Q - 2\alpha$ . Also, it is easy to check the shape-invariance condition

$$V_+(r, \rho_0) = V_-(r, \rho_i) + R(\rho_i) \qquad (29)$$

which holds via the mapping $Q \to Q - 2\alpha$ .In this study $\rho_0 = Q$ and $\rho_i$ is a function of $\rho_0$, i.e , $\rho_1 = f(\rho_0) = \rho_0 - 2\alpha$ . Thus, $\rho_n = \rho_0 - 2\alpha n$ and from Eq.(29), we write

$$R(\rho_n) = \left(\frac{-\rho_{n-1}^2 + \omega_1 - \omega_3 q^2}{2q\rho_{n-1}}\right)^2 - \left(\frac{-\rho_n^2 + \omega_1 - \omega_3 q^2}{2q\rho_n}\right)^2 \qquad (30)$$

$$\tilde{E}_{n,l}^{-} = \sum_{k=1}^{n} R(\rho_k) = \left(\frac{-\rho_0^2 + \omega_1 - \omega_3 q^2}{2q\rho_0}\right)^2 - \left(\frac{-\rho_n^2 + \omega_1 - \omega_3 q^2}{2q\rho_n}\right)^2 \qquad (31)$$

Using Eqs. (15), (26) and (30), we obtained the transcendental energy spectrum for the Hulthen plus deformed hyperbolic potential model for the Klein-Gordon equation in D-dimension as follows,

$$\tilde{E}_{n,l} = \tilde{E}_{n,l}^{-} + \tilde{E}_{0,l} = -\left(\frac{-\rho_n^2 + \omega_1 - \omega_3 q^2}{2q\rho_n}\right)^2 + \omega_3 \qquad (32)$$

More explicitly, we write the energy equation (32) as,

$$E_{nJ_D}^2 - m_0^2 = -\frac{\hbar^2 c^2}{4q^2}\left(\frac{\omega_3 q^2 - \omega_1}{2\alpha(n+\sigma)} + 2\alpha(n+\sigma)\right)^2 + \hbar^2 c^2 \omega_3 \qquad (33)$$

Where

$$\sigma = \frac{q}{2}\left(1 + \sqrt{1 + \frac{1}{\alpha^2 q^2}\left(\omega_1 + \omega_2 q + \omega_3 q^2\right)}\right) \qquad (34)$$

In order to determine the corresponding wave function for the system, we make a change of variable in Eq. (14) by writing $z = -qe^{-2\alpha r}$ to obtain

$$\frac{d^2 U_{nJ_D}}{dz^2} + \frac{1}{z}\frac{dU_{nJ_D}}{dz} + \frac{1}{z^2(1-z)^2}\left(-\chi_1 z^2 + \chi_2 z + \chi_3\right)U_{nJ_D}(z) = 0\,, \qquad (32)$$

Where

$$\chi_1 = \frac{1}{4\alpha^2 q^2 \hbar^2 c^2}\left[2m_0 c^2 S_0 q - 2m_0 c^2 S_1 q^2 + 2E_{n,J_D} V_0 q - 2E_{n,J_D} V_1 q^2 + S_0^2 + S_1^2 q^2 - V_0^2 + V_1^2 q^2\right] - \frac{E_{n,J_D}^2 - m^2 c^4}{4\alpha^2 \hbar^2 c^2}$$

$$\chi_2 = -\frac{1}{4\alpha^2 q \hbar^2 c^2}\begin{bmatrix} -2m_0 c^2 S_0 - 2m_1 c^2 S_0 + 2m_1 c^2 S_1 q - 2E_{nJ_D} V_0 - 2S_0 S_1 q + 2S_1^2 q + 2V_0 V_1 q + \\ 2V_1^2 q - 2m_0 m_1 c^4 q + 4\gamma \hbar^2 c^2 \alpha^2 \end{bmatrix} - \frac{2(E_{n,J_D}^2 - m^2 c^4)}{4\alpha^2 \hbar^2 c^2}$$

$$\chi_3 = \frac{1}{4\alpha^2 \hbar^2 c^2}\left[2m_1 c^2 S_1 + 2m_0 c^2 S_1 + 2E_{nJ_D} V_1 - 2S_0 S_1 + S_1^2 + 2V_0 V_1 + V_1^2 + 2m_0 m_2 c^4 + m_1^2 c^4\right] - \frac{(E_{n,J_D}^2 - m^2 c^4)}{4\alpha^2 \hbar^2 c^2} \qquad (33)$$

The solution of Eq. (32) becomes

$$U_{nJ_D}(r) = N_{nJ_D}\left(qe^{-2\alpha r}\right)^{\sqrt{\chi_3}}\left(1-qe^{-2\alpha r}\right)^{\sqrt{\frac{1}{4}+\chi_1-\chi_2+\chi_3}} P_n^{\left(2\sqrt{\chi_3},2\sqrt{\frac{1}{4}+\chi_1-\chi_2+\chi_3}\right)}\left(1-2qe^{-2\alpha r}\right) \quad (34)$$

where $N_{n,l}$ is the normalization constant and $P_n^{(\mu,\nu)}(x)$ is the Jacobi polynomial.

**3 Scattering state solutions**

With the change of variable $x = 1-z$ in Eq.(32), we get

$$x(1-x)U''(x) - xU'(x) + \left(\frac{\Xi_1}{x} + \frac{\Xi_2}{1-x} + \Xi_3\right)U(x) = 0 \quad (37)$$

Where,

$$\begin{aligned} \Xi_1 &= \chi_2 + \chi_3 - \chi_1 \\ \Xi_2 &= \chi_3 \\ \Xi_3 &= \chi_1 \end{aligned} \quad (38)$$

and with transformation $F_{n,l} = y^{\lambda_1}\left(1-y\right)^{\lambda_2}\varphi(z)$ in Eq.(37), we obtain the hypergeometric function in the form[34]

$$z(1-z)\varphi''(z) + \left(1+2\eta_1 - (1+2\eta_1+2\eta_2)z\right)\varphi(z) - \eta_1\eta_2\varphi(z) = 0 \quad (39)$$

Where,

$$\begin{aligned} \lambda_1 &= \frac{1}{2}\left(1+\sqrt{1-4\Xi_1}\right), \\ \lambda_2 &= -\frac{ik}{2\alpha}, \quad k = \sqrt{4\alpha^2\Xi_2} \end{aligned} \quad (40)$$

The solutions of Eq.(39) is the hypergeometric function given by

$$\varphi(y) = {}_2F_1\left(\eta_1,\eta_2,\eta_3,y\right) \quad (41)$$

Where,

$$\begin{aligned} \eta_1 &= \frac{1}{2}\left(1+\sqrt{1-4\Xi_1}\right) - \frac{ik}{\alpha} + \sqrt{\Xi_3}, \\ \eta_2 &= \frac{1}{2}\left(1+\sqrt{1-4\Xi_1}\right) - \frac{ik}{\alpha} - \sqrt{\Xi_3}, \\ \eta_3 &= 1+\sqrt{1-4\Xi_1}, \end{aligned} \quad (42)$$

We can write the complete wave function as,

$$F_{n,l}(r) = N_{n,l}\left(-qe^{-2\alpha r}\right)^{\lambda_1}\left(1-qe^{-2\alpha r}\right)^{\lambda_2} {}_2F_1\left(\eta_1,\eta_2,\eta_3;1-qe^{-2\alpha r}\right) \quad (43)$$

The scattering state for $E > 0$, in 3D case the boundary conditions for the scattering state are defined as follows [51-52],

$$U(r) \to 0,$$
$$r \to 0$$
$$U(r) \to 2\sin\left[kr + \delta - \frac{l_D \pi}{2}\right], \qquad (44)$$
$$r \to \infty$$

However, the extension to the D-dimensional case takes the form [51-52],

$$U(r) \to 0,$$
$$r \to 0$$
$$U(r) \to 2\sin\left[kr + \delta - \frac{\pi}{2}\left(l_D + \frac{(D-3)}{2}\right)\right], \qquad (45)$$
$$r \to \infty$$

Eq.(45) ensures us that the radial wave functions of the scattering state for exponential type potential are also normalized, where $\delta_{l_D}$ is the phase shift in D-dimensions.

Now we study the asymptotic form of Eq.(43) for large values of $r$ and calculate the normalization constant $N_{nl_D}$ and the phase shift. From Eq.(42),we get

$$\eta_3 - \eta_1 - \eta_2 = \frac{2ik}{\alpha} = \left(\eta_1 + \eta_2 - \eta_3\right)^* \qquad (46)$$

$$\eta_3 - \eta_1 = \frac{1}{2}\left(1 + \sqrt{1 - 4\Xi_1}\right) + \frac{ik}{\alpha} - \sqrt{\Xi_3} = \eta_2^* \qquad (47)$$

$$\eta_3 - \eta_2 = \frac{1}{2}\left(1 + \sqrt{1 - 4\Xi_1}\right) - \frac{ik}{\alpha} + \sqrt{\Xi_3} = \eta_1^* \qquad (48)$$

Now applying the transformation properties for the hypergeometric function [53],

$${}_2F_1\left(\eta_1, \eta_2, \eta_3; 0\right) = 1, \qquad (49)$$

$${}_2F_1\left(\eta_1, \eta_2, \eta_3; z\right) = \frac{\Gamma(\eta_3)\Gamma\left(\eta_3 - \eta_1 - \eta_2\right)}{\Gamma\left(\eta_3 - \eta_1\right)\Gamma\left(\eta_3 - \eta_2\right)}\, {}_2F_1\left(\eta_3; \eta_2; \eta_1 + \eta_2 - \eta_3 + 1; 1 - z\right)$$
$$+ \left(1 - z\right)^{\eta_3 - \eta_1 - \eta_2} \frac{\Gamma(\eta_3)\Gamma\left(\eta_1 + \eta_2 - \eta_3\right)}{\Gamma(\eta_1)\Gamma(\eta_2)}\, {}_2F_1\left(\eta_3 - \eta_1; \eta_3 - \eta_2; \eta_3 - \eta_1 - \eta_2 + 1; 1 - z\right) \qquad (50)$$

Thus, the term ${}_2F_1\left(\eta_1, \eta_2, \eta_3; 1 - qe^{-2\alpha r}\right)$ in Eq.(43) as $r \to \infty$ becomes,

$$
{}_2F_1\left(\eta_1,\eta_2,\eta_3;1-qe^{-2\alpha r}\right)=\Gamma(\eta_3)\left[\frac{\Gamma\left(\eta_3-\eta_1-\eta_2\right)}{\Gamma\left(\eta_3-\eta_1\right)\Gamma\left(\eta_3-\eta_2\right)}+\left[\frac{\Gamma\left(\eta_3-\eta_1-\eta_2\right)}{\Gamma((\eta_3-\eta_1))\Gamma(\eta_3-\eta_2)}\right]^* q^{(\eta_3-\eta_1-\eta_2)}e^{-\alpha(\eta_3-\eta_1-\eta_2)r}\right],
$$

$$
=\Gamma(\eta_3)q^{i(\eta_3-\eta_1-\eta_2)}\left[\frac{\Gamma\left(\eta_3-\eta_1-\eta_2\right)q^{-i(\eta_3-\eta_1-\eta_2)}}{\Gamma\left(\eta_3-\eta_1\right)\Gamma\left(\eta_3-\eta_2\right)}+\left[\frac{\Gamma\left(\eta_3-\eta_1-\eta_2\right)q^{-i(\eta_3-\eta_1-\eta_2)}}{\Gamma(\eta_3-\eta_1)\Gamma(\eta_3-\eta_2)}\right]^* e^{-i\alpha(\eta_3-\eta_1-\eta_2)r}\right] \quad (51)
$$

We can simplify Eq. (51) as follows,

$$
{}_2F_1\left(\eta_1,\eta_2,\eta_3;1-qe^{-2\alpha r}\right)=\Gamma(\eta_3)q^{\frac{ik}{\alpha}}\left[\frac{\Gamma\left(\eta_3-\eta_1-\eta_2\right)q^{-\frac{ik}{\alpha}}}{\Gamma\left(\eta_3-\eta_1\right)\Gamma\left(\eta_3-\eta_2\right)}+\left[\frac{\Gamma\left(\eta_3-\eta_1-\eta_2\right)q^{-\frac{ik}{\alpha}}}{\Gamma(\eta_3-\eta_1)\Gamma(\eta_3-\eta_2)}\right]^* e^{-ikr}\right]
$$

$$
=\Gamma(\eta_3)q^{\frac{ik}{\alpha}}\left[\frac{\Gamma\left(2ik/\alpha\right)q^{-\frac{ik}{\alpha}}}{\Gamma\left(\eta_2^*\right)\Gamma\left(\eta_1^*\right)}+\left[\frac{\Gamma\left(2ik/\alpha\right)q^{-\frac{ik}{\alpha}}}{\Gamma(\eta_2^*)\Gamma(\eta_1^*)}\right]^* e^{-ikr}\right] \quad (52)
$$

Also using the relations

$$
\frac{\Gamma\left(\eta_3-\eta_1-\eta_2\right)}{\Gamma\left(\eta_3-\eta_1\right)\Gamma\left(\eta_3-\eta_2\right)}=\left|\frac{\Gamma\left(\eta_3-\eta_1-\eta_2\right)}{\Gamma\left(\eta_3-\eta_1\right)\Gamma\left(\eta_3-\eta_2\right)}\right|e^{i\delta},
$$

$$
=\left|\frac{\Gamma\left(2ik/\alpha\right)}{\Gamma\left(\eta_2^*\right)\Gamma\left(\eta_1^*\right)}\right|e^{i\delta} \quad (53)
$$

$$
q^{-\frac{ik}{\alpha}}=\left|q^{-\frac{ik}{\alpha}}\right|e^{i\delta'} \quad (54)
$$

and inserting it into Eq. (51), we have

$$
{}_2F_1\left(\eta_1,\eta_2,\eta_3;1-qe^{-2\alpha r}\right)=\Gamma(\eta_3)q^{\frac{ik}{\alpha}}\left|\frac{\Gamma\left(2ik/\alpha\right)q^{-\frac{ik}{\alpha}}}{\Gamma\left(\eta_2^*\right)\Gamma\left(\eta_1^*\right)}\right|\left[e^{i(\delta+\delta')}+e^{-2ikr}e^{-i(\delta+\delta')}\right]
$$

$$
=\Gamma(\eta_3)q^{\frac{ik}{\alpha}}\left|\frac{\Gamma\left(2ik/\alpha\right)q^{-\frac{ik}{\alpha}}}{\Gamma\left(\eta_2^*\right)\Gamma\left(\eta_1^*\right)}\right|e^{-ikr}\left[e^{i\left(\delta+\delta'-\frac{k\ell n2}{\alpha}+ikr\right)}+e^{-i\left(\delta+\delta'-\frac{k\ell n2}{\alpha}+ikr\right)}\right]
$$

$$
=\Gamma(\eta_3)q^{\frac{ik}{\alpha}}\left|\frac{\Gamma\left(2ik/\alpha\right)q^{\frac{ik}{\alpha}}}{\Gamma\left(\eta_2^*\right)\Gamma\left(\eta_1^*\right)}\right|e^{-ikr}\sin\left(kr+\delta+\delta'-\frac{k\ell n2}{\alpha}+\frac{\pi}{2}\left[l_D+\frac{(D-3)}{2}\right]+\frac{\pi}{2}\left[l_D+\frac{(D-1)}{2}\right]\right) \quad (55)
$$

Therefore, the asymptotic form of Eq. (43) for $r \to \infty$ becomes

$$F_{n,l_D} = 2N_{n,l_D}\Gamma(\eta_3)\left|\frac{\Gamma\left(2ik/\alpha\right)q^{\frac{ik}{\alpha}}}{\Gamma\left(\eta_2^*\right)\Gamma\left(\eta_1^*\right)}\right| e^{-ikr}\sin\left(kr+\delta+\delta'-\frac{k\ell n2}{\alpha}+\frac{\pi}{2}\left[l_D+\frac{(D-3)}{2}\right]+\frac{\pi}{2}\left[l_D+\frac{(D-1)}{2}\right]\right) \quad (56)$$

Now comparing Eq.(56) with the boundary conditions[54-56], $r \to \infty \Rightarrow F(\infty) \to 2\sin\left(kr+\delta_{l_D}-\frac{\pi}{2}\left(l_D+\frac{(D-3)}{2}\right)\right)$, we get the phase shift and the normalization constant as follows:

$$\begin{aligned}\delta_{l_D} &= \frac{\pi}{2}\left(l_D+\frac{(D-3)}{2}\right)+\delta+\delta' \\ &= \frac{\pi}{2}\left[l_D+\frac{(D-1)}{2}\right]-\frac{k\ell n2}{\alpha}+\arg\Gamma\left(\eta_3-\eta_2-\eta_1\right)-\arg\Gamma\left(\eta_3-\eta_1\right)-\arg\Gamma\left(\eta_3-\eta_2\right)+\arg\left(q^{\frac{-ik}{\alpha}}\right)\end{aligned} \quad (57)$$

$$N_{n,l_D} = \frac{1}{\Gamma(\eta_3)}\left|\frac{\Gamma\left(\eta_2^*\right)\Gamma\left(\eta_1^*\right)}{\Gamma\left(2ik/\alpha\right)q^{\frac{ik}{\alpha}}}\right| \quad (58)$$

It had been shown that the poles of the S-matrix in the complex energy plane determined the bound states for the real poles and the scattering states for the complex poles in the lower half of the energy plane [57].In order to obtain the bound state solutions, we used the following definition of the gamma functions,

$$\Gamma\left(z\right) = \frac{\Gamma\left(z+1\right)}{z} = \frac{\Gamma\left(z+2\right)}{z(z+1)} = \frac{\Gamma\left(z+3\right)}{z(z+1)(z+2)} = \ldots \quad (59)$$

Where the $z = 0,-1,-2\ldots$ are the first order poles of the gamma function $\Gamma(z)$.Thus, the first order poles in Eq.(59) is $\Gamma\left(\eta_3-\eta_1\right)$ and satisfy the condition,

$$\frac{1}{2}\left(1+\sqrt{1-4\Xi_1}\right)-\frac{ik}{\alpha}-\sqrt{\Xi_3} = -n_r \quad (60)$$

where, $n_r = 0,1,2\ldots$ At the poles of the scattering amplitude the bound state energy level in D-dimension is given by equation (60) and is the same as equation (33).

## 4 Results and Discussion

In order to test the accuracy of our work, we computed numerically the bound state energy for the Hulthen plus deformed hyperbolic potential for arbitrary $l$-state in D-dimension in Table 1 and we also show the plot of the behaviour of the energy level as a function of the potential parameters in fig.4-11 We also examine the special cases of the phase shift factor and normalization constant and compare our results with the available literature.

### 4.1 Scattering state solutions of generalized Hulthen potential in D-dimensions

If we choose, $V_1 = 0, S_0 = S_1 = m_1 = 0$ and $\alpha \to \frac{\alpha}{2}$ in Eq.(2),we obtain the generalized Hulthen potential as,

$$V_H(r) = -V_0 \frac{e^{-\alpha r}}{1 - qe^{-\alpha r}} \qquad (61)$$

If we substitute these parameters into Eqs.,(57) and (58) we obtain the phase factor and normalization constant for the Hulthen potential in D-dimension as,

$$\delta_J^H = \frac{\pi}{2}\left[J_D + \frac{(D-1)}{2}\right] - \frac{k\ell n2}{\alpha} + \arg\Gamma\left(\frac{2ik}{\alpha}\right) - \arg\Gamma\left(\lambda_1 - \frac{ik}{\alpha} - \sqrt{\Xi_3}\right)$$
$$-\arg\Gamma\left(\lambda_1 - \frac{ik}{\alpha} - \sqrt{\Xi_3}\right) + \arg\left(q^{\frac{-ik}{\alpha}}\right) \qquad (62)$$

$$N_{nJ_D}^H = \frac{\Gamma\left(\lambda_1^H - \frac{ik}{\alpha} - \sqrt{\Xi_3}\right)}{\Gamma(1 + 2\lambda_1^H)} \times \frac{\Gamma\left(\lambda_1^H - \frac{ik}{\alpha} - \sqrt{\Xi_3}\right)}{\Gamma\left(2ik/\alpha\right) q^{\frac{ik}{\alpha}}} \qquad (63)$$

These results are consistent with that reported by Feng et al [51] when $D = 3$ ..The normalized wave function for the Hulthen potential becomes,

$$U_{nJ_D}^H(r) = N_{nJ_D}^H \left(1 - qe^{-\alpha r}\right)^{\lambda_1^H} (q)^{\frac{ik}{\alpha}} e^{-ikr} {}_2F_1\left(\eta_1^H, \eta_2^H, \eta_3^H, 1 - qe^{-\alpha r}\right) \qquad (64)$$

If $q = 1$ then the generalized Hulthen potential reduced to the standard potential reported by Chen et al [57].

### 4.2 Scattering state solutions of Woods-Saxon potential in D-dimensions

Again if we set $\alpha = \frac{1}{R}, q = e^{\frac{\theta}{R}}$ and $V_0 \to V_0 q = V_0 e^{\frac{\theta}{R}}$ as given in Ref.[51],the generalized Hulthen potential reduces to the standard Woods-Saxon potential [58],

$$V(r) = \frac{V_0}{1 - \exp\left[\left(\frac{r-\theta}{R}\right)\right]} \qquad (65)$$

where $V_0$ is the potential depth, $\theta$ is the width of the potential and $R$ is the surface thickness whose values correspond to the ionization energies. We obtain the corresponding phase factor and normalization for the Woods-Saxon potential in d-dimensions as follows,

$$\delta_J^{WS} = \frac{\pi}{2}\left[J_D + \frac{(D-1)}{2}\right] - kR\ell n2 + \arg\Gamma\left(2ikR\right) - \arg\Gamma\left(\lambda_1^{WS} - iRk + \sqrt{\Xi_3}\right)$$
$$-\arg\Gamma\left(\lambda_1^{WS} + ikR - \sqrt{\Xi_3}\right) + \arg\left(\left(e^{-\frac{\theta}{R}}\right)^{-iRk}\right) \qquad (66)$$

$$N_{nJ_D}^{WS} = \frac{\Gamma\left(\lambda_1^{WS} - ikR - \sqrt{\Xi_3}\right)}{\Gamma(1+2\lambda_1^{WS})} \times \frac{\Gamma\left(\lambda_1^{WS} + ikR - \sqrt{\Xi_3}\right)}{\Gamma\left(2ikR\right)\left(e^{\frac{\theta}{R}}\right)^{ikR}} \qquad (67)$$

$$U_{nJ_D}^{WS}(r) = N_{nJ_D}^{WS}\left(1 - \exp\left(\frac{\theta - r}{R}\right)\right)^{\mu^{WS}}\left(\exp\left(\frac{\theta}{r}\right)\right)^{ikR} e^{-ikr}\, {}_2F_1\left(a,b,c,1-\exp\left(\frac{(\theta - r)}{R}\right)\right).$$

The above result is the same as the one reported by Feng et al [51] when $D = 3$.

**5 Conclusions**

In this paper, we have studied the Klein-Gordon equation with Hulthen plus deformed hyperbolic potential and obtain the bound state energy eigenvalues and the scattering state phase factor. Special cases of the potential are discussed in details. The present findings may have many applications in different branches of physics and quantum chemistry.

Table 1, Energy of the system for different states and dimensions
$m_0 = -5, m_1 = -0.2, V_0 = 2$ , $\alpha = 0.01, S_1 = 3, V_1 = 0.5, q = 1$

| | $l$ | $E_{n,l}^D(D=1)$ | $E_{n,l}^D(D=2)$ | $E_{n,l}^D(D=3)$ | $E_{n,l}^D(D=4)$ |
|---|---|---|---|---|---|
| $n=0$ | 0 | 2.569172676<br>-2.203041293 | 2.569156234<br>-2.203011271 | 2.569172676<br>-2.203041203 | 2.569221997<br>-2.203130998 |
| | 1 | 2.569172676<br>-2.203041203 | 2.569221997<br>-2.203130998 | 2.569304187<br>-2.203280640 | 2.569419235<br>-2.203490115 |
| | 2 | 2.569304187<br>-2.203280640 | 2.569419235<br>-2.203490115 | 2.569567123<br>-2.203759397 | 2.569747824<br>-2.204088452 |
| | 3 | 2.569567123<br>-2.203759397 | 2.569747824<br>-2.204088452 | 2.569961309<br>-2.204477240 | 2.570207541<br>-2.204925711 |
| | 4 | 2.569961309<br>-2.204477240 | 2.570207541<br>-2.204925711 | 2.570486483<br>-2.205433819 | 2.570798087<br>-2.206001495 |
| $n=1$ | 0 | 2.5953363031<br>-2.251003944 | 2.595347227<br>-2.250974832 | 2.595363031<br>-2.251003944 | 2.595410441<br>-2.251091282 |
| | 1 | 2.595363031<br>-2.251003944 | 2.595410441<br>-2.251091282 | 2.595489446<br>-2.251236828 | 2.595600036<br>-2.251440570 |
| | 2 | 2.595489446<br>-2.251236828 | 2.595600036<br>-2.251440570 | 2.595742192<br>-2.251702481 | 2.595915893<br>-2.252022535 |
| | 3 | 2.595742192<br>-2.251702481 | 2.595915893<br>-2.252022535 | 2.596121109<br>-2.252400688 | 2.596357804<br>-2.252836894 |
| | 4 | 2.596121109<br>-2.252400688 | 2.596357804<br>-2.252836894 | 2.596625945<br>-2.253331107 | 2.596925485<br>-2.253883264 |
| $n=2$ | 0 | 2.620547699<br>-2.297667736 | 2.620532497<br>-2.297639403 | 2.620547699<br>-2.297667736 | 2.620593305<br>-2.297752733 |
| | 1 | 2.620547699<br>-2.297667736 | 2.620593305<br>-2.297752733 | 2.620669304<br>-2.297894378 | 2.620775688<br>-2.2980922663 |
| | 2 | 2.620669304<br>-2.297894378 | 2.620775688<br>-2.298092663 | 2.620912436<br>-2.298347556 | 2.621079530<br>-2.298659035 |
| | 3 | 2.620912436<br>-2.298347556 | 2.621079530<br>-2.298659035 | 2.621276940<br>-2.299027058 | 2.621504637<br>-2.299451584 |
| | 4 | 2.621276940<br>-2.299027058 | 2.621504637<br>-2.299451584 | 2.621762582<br>-2.299932564 | 2.622050736<br>-2.300469940 |

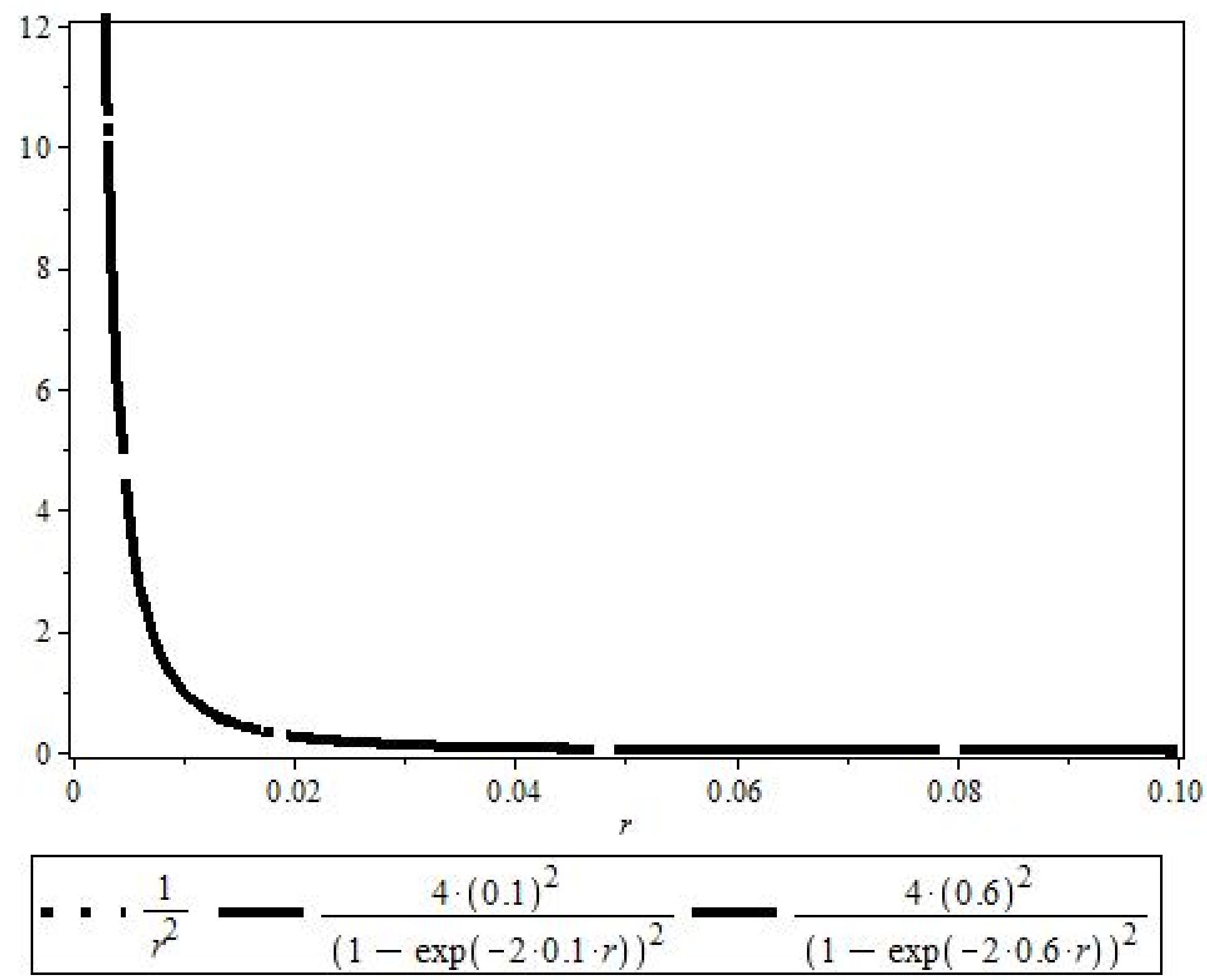

Fig.3: $\frac{1}{r^2}$ **and its approximation of equation (1)**

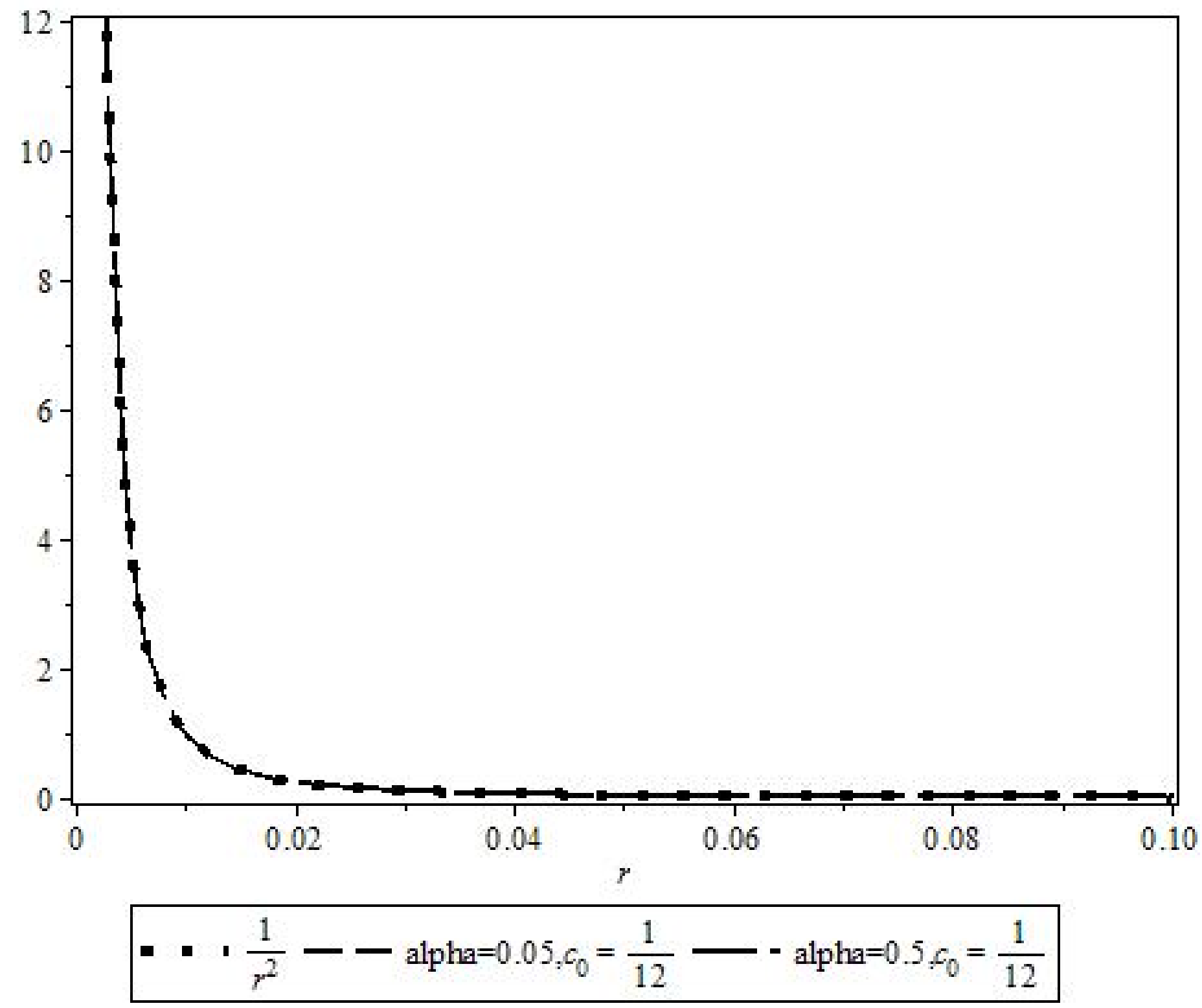

Fig.1: $\frac{1}{r^2}$ **and its approximation of equation (13)**

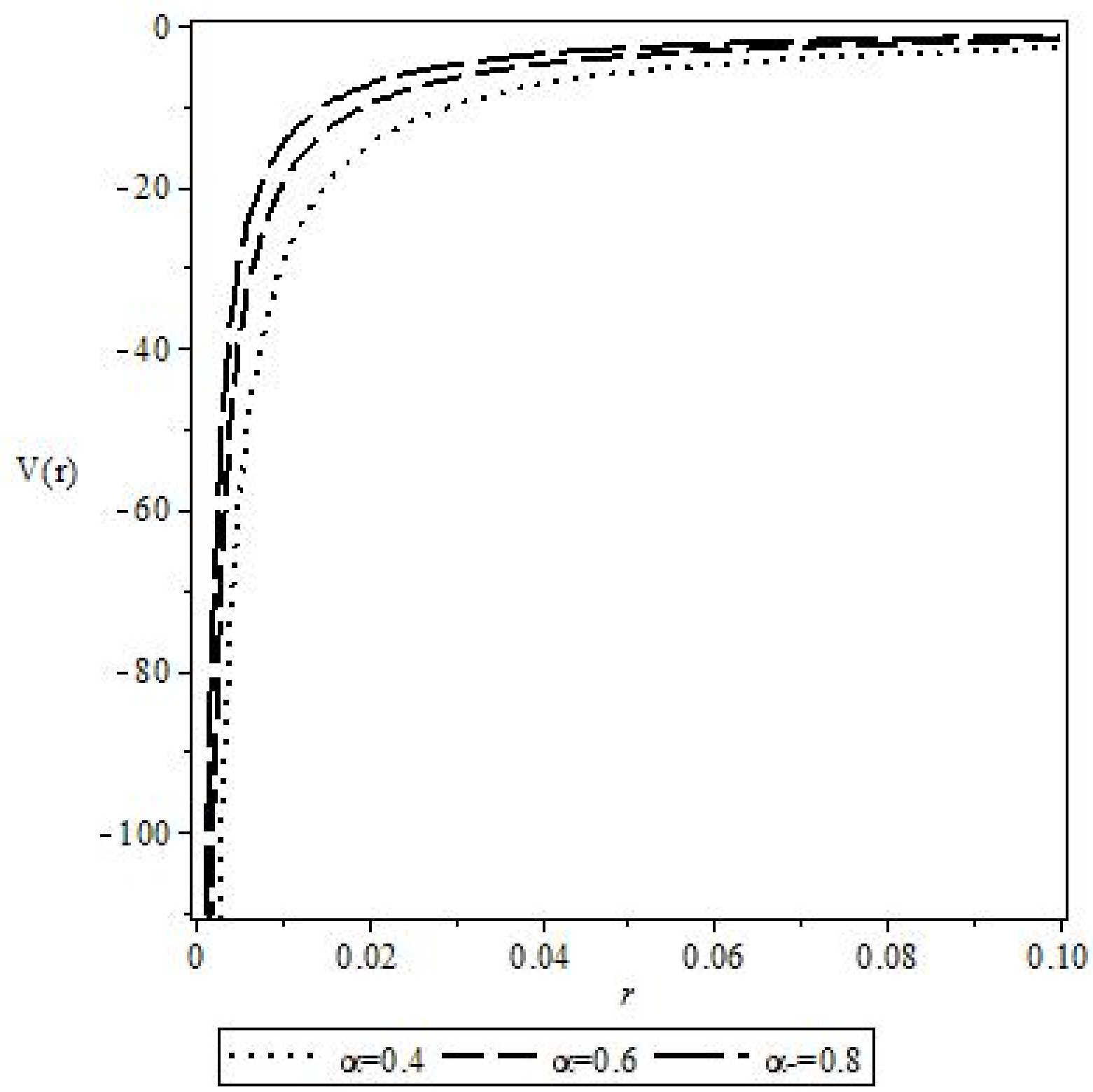

Fig.2: Shape of the Hulthen plus deformed hyperbolic potential for $V_0 = 0.2\, fm^{-1}, V_1 = 0.04\, fm^{-1}$ and various values of $\alpha = 0.4, 0.6$ and $0.8$ respectively.

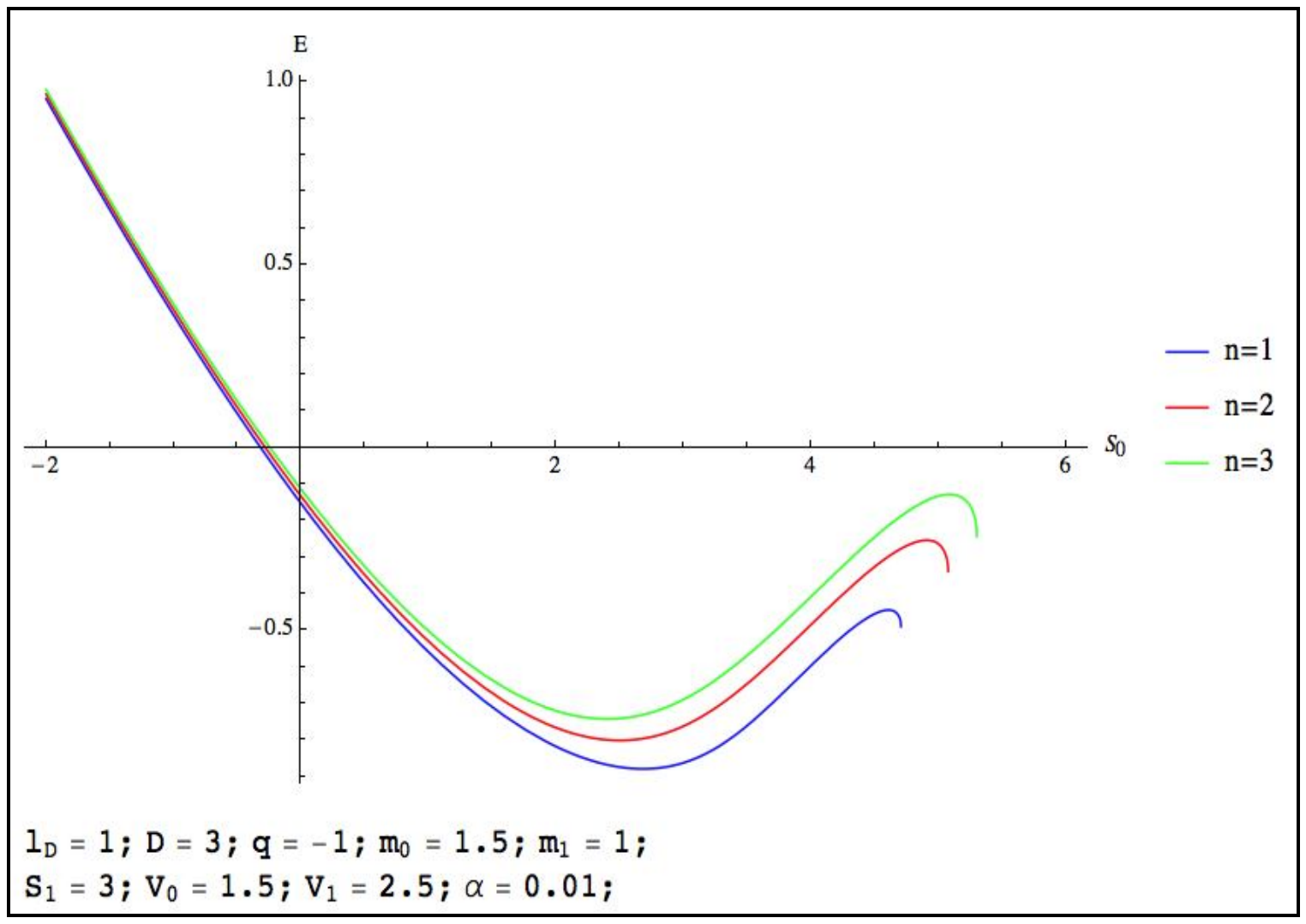

Fig.4: Behaviour of the energy level of the system versus $S_0$ for various values of $n = 1, 2$ and $3$ respectively.

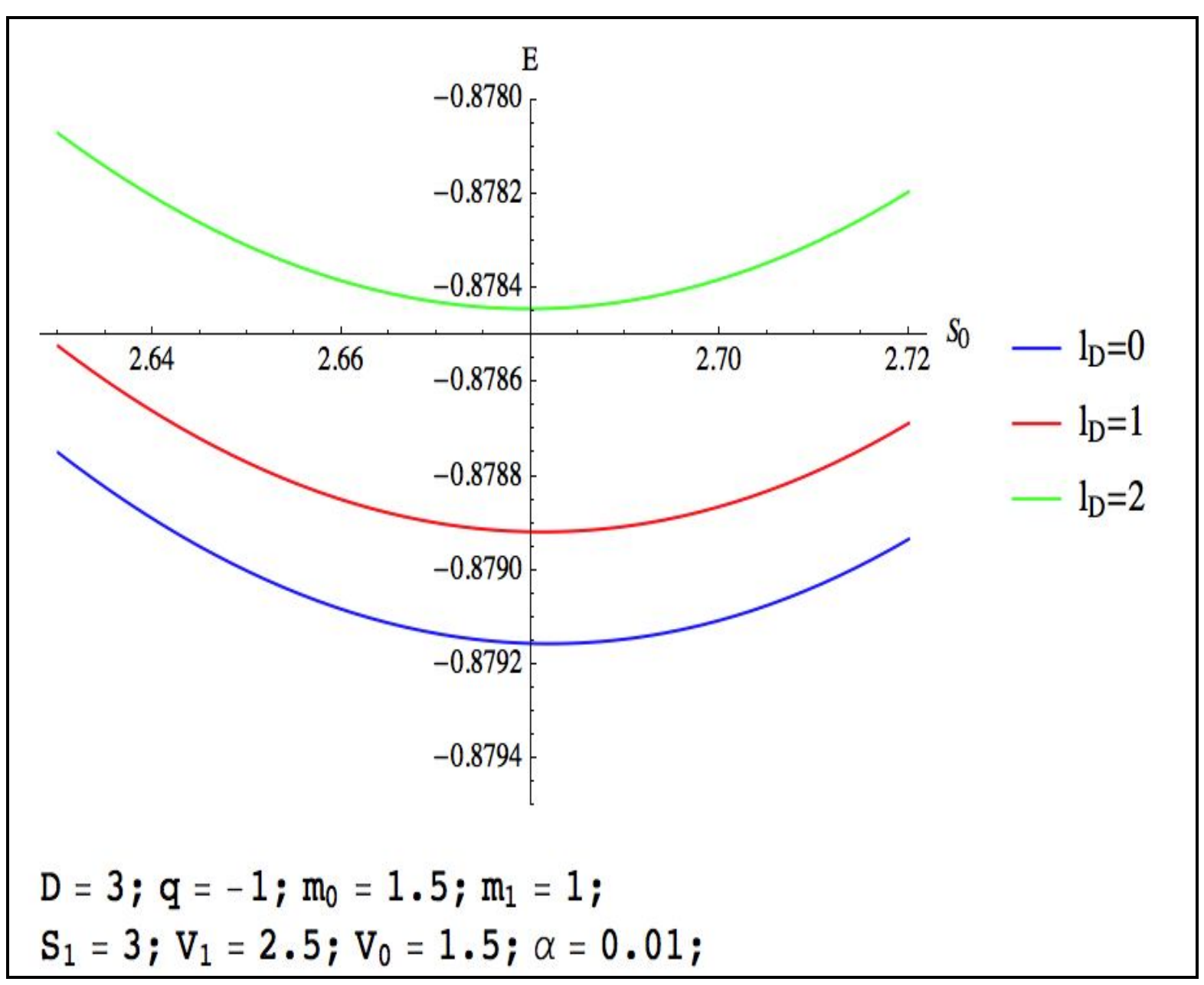

Fig.5: Variation of the energy spectrum with $S_0$ for different values of the angular momentum quantum number $l_D = 0,1$ and $2$ respectively.

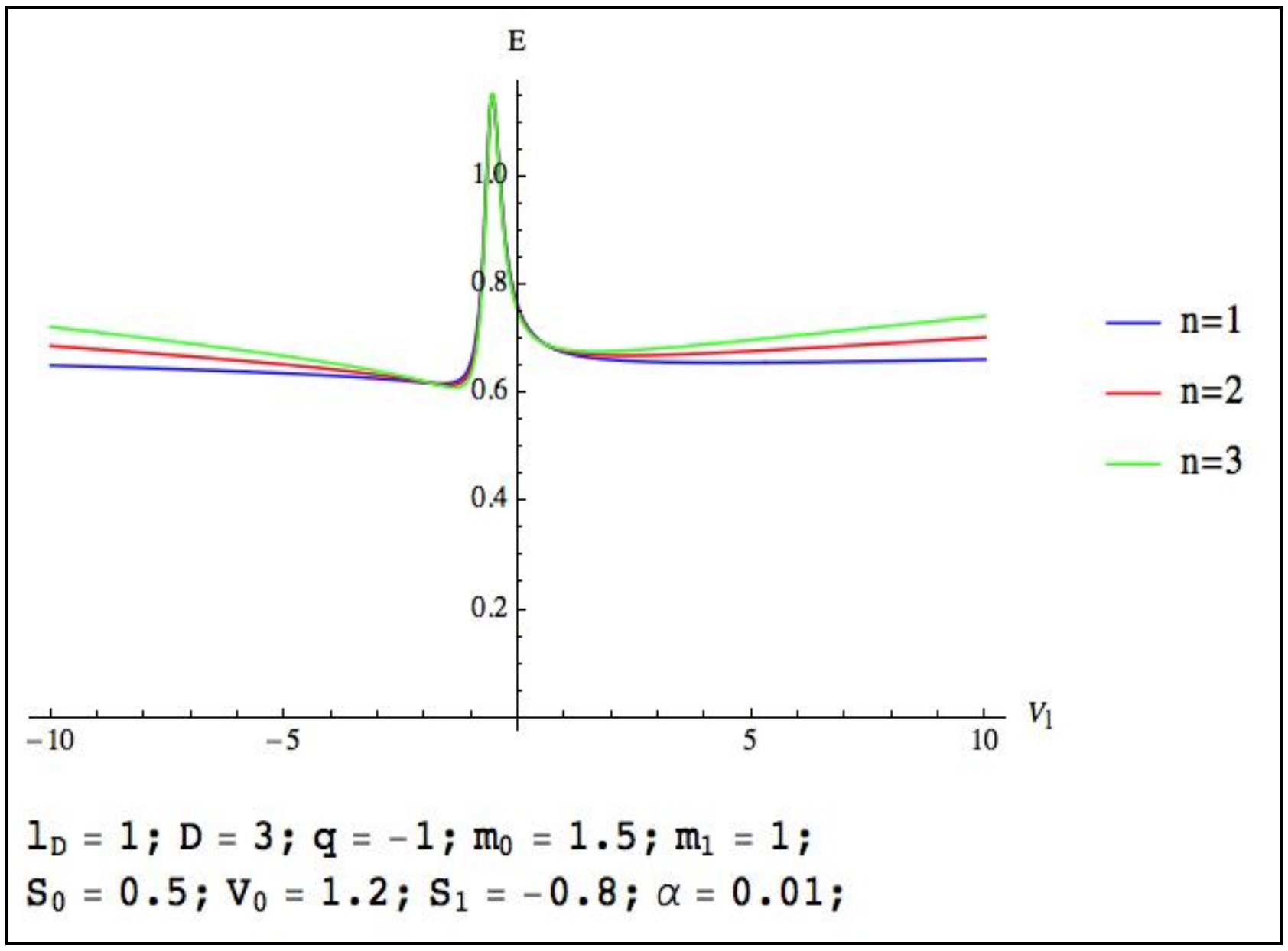

Fig.6: Plot of the energy level versus $V_1$ for various values of the principal quantum number $n = 1, 2$ and 3 respectively.

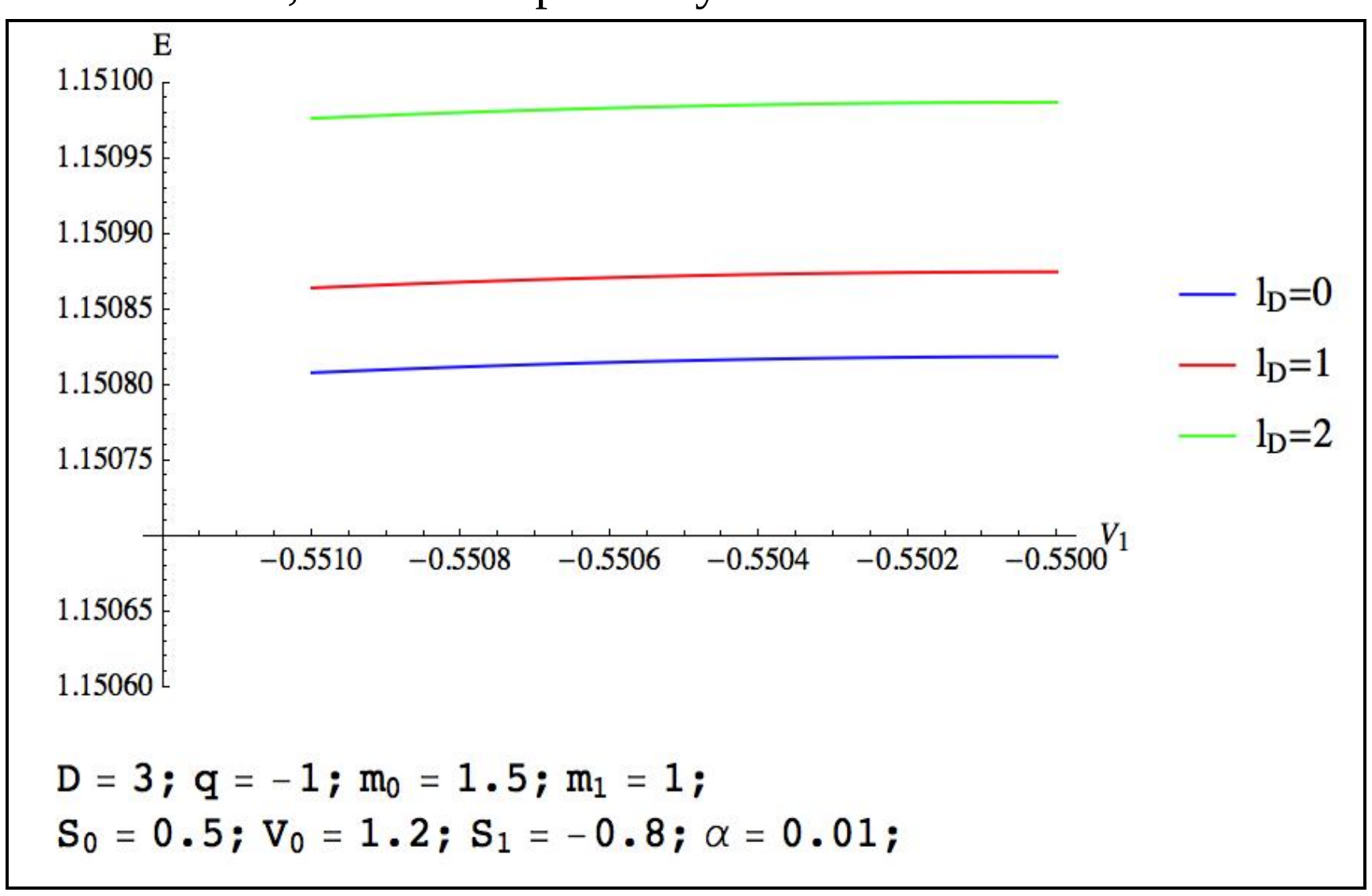

Fig.7: Variation of the energy eigenvalues with $V_1$ for different values of $l_D = 0,1$ and 2 respectively.

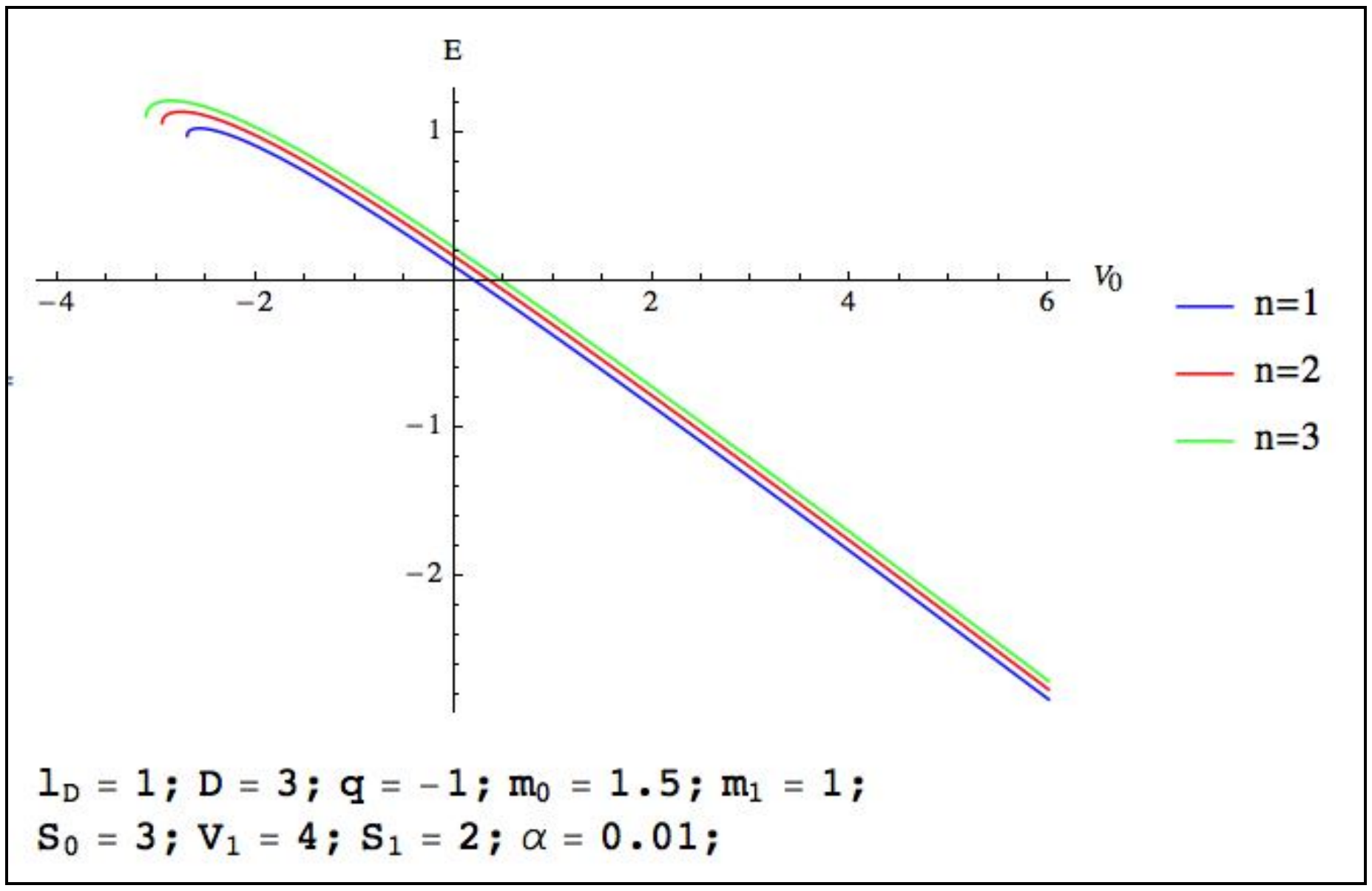

Fig.8: Energy variation with $V_0$ for $n = 1, 2$ and 3 respectively.

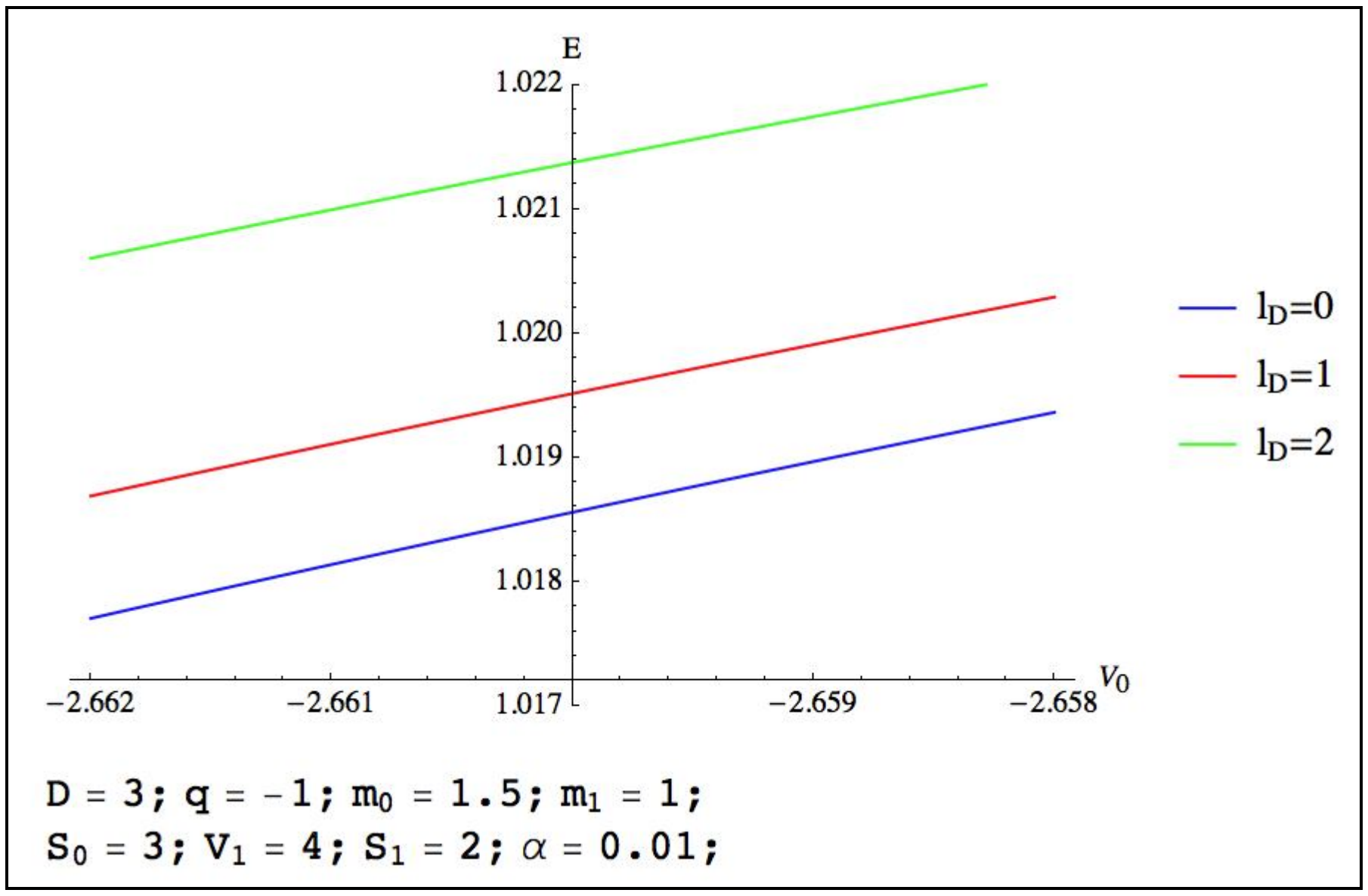

Fig.9: Energy changes with $V_0$ for different values of $l_D = 0,1$ and 2 respectively.

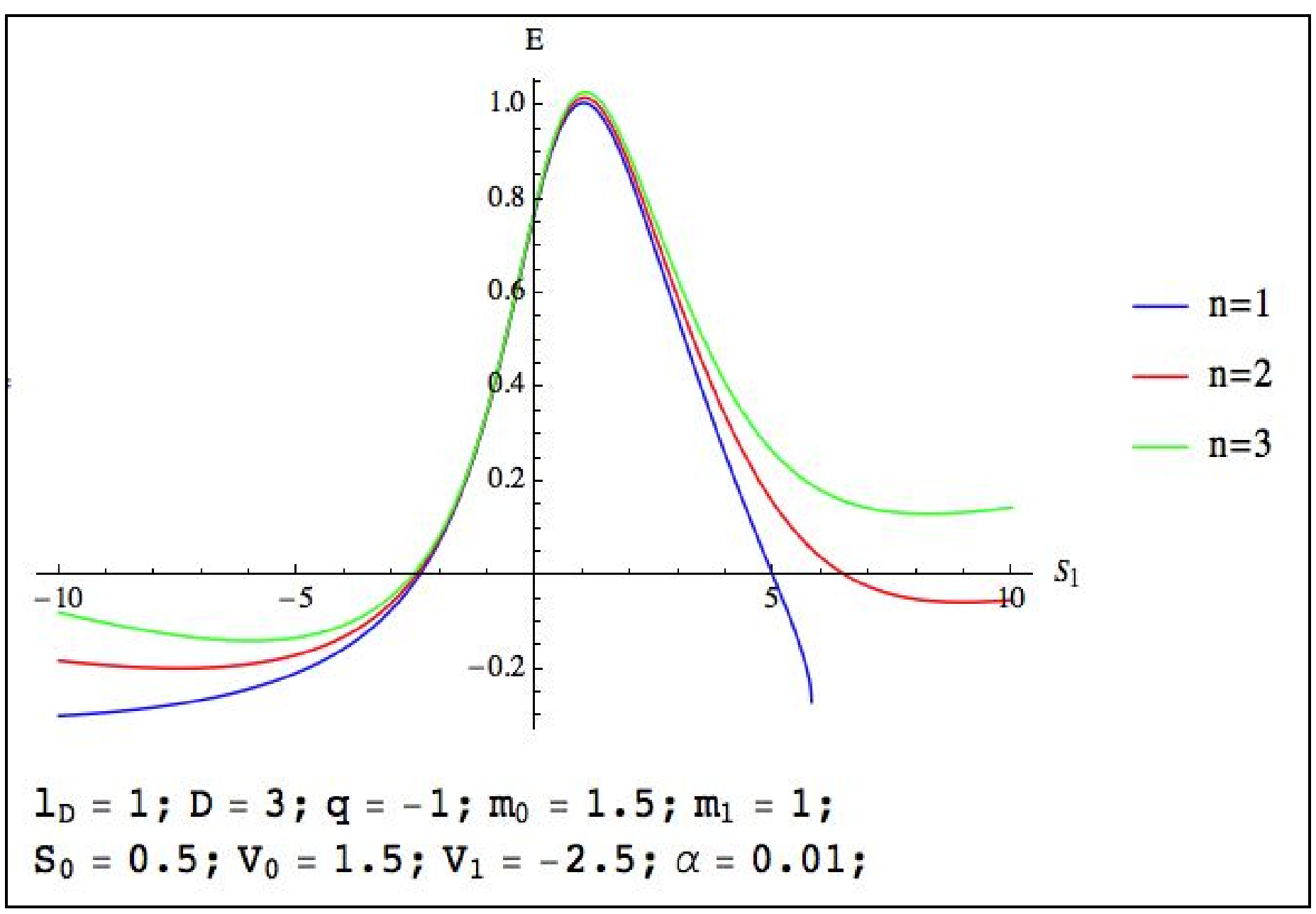

Fig.10: Plot of the energy spectrum with $S_1$ for $n = 1, 2$ and 3 respectively.

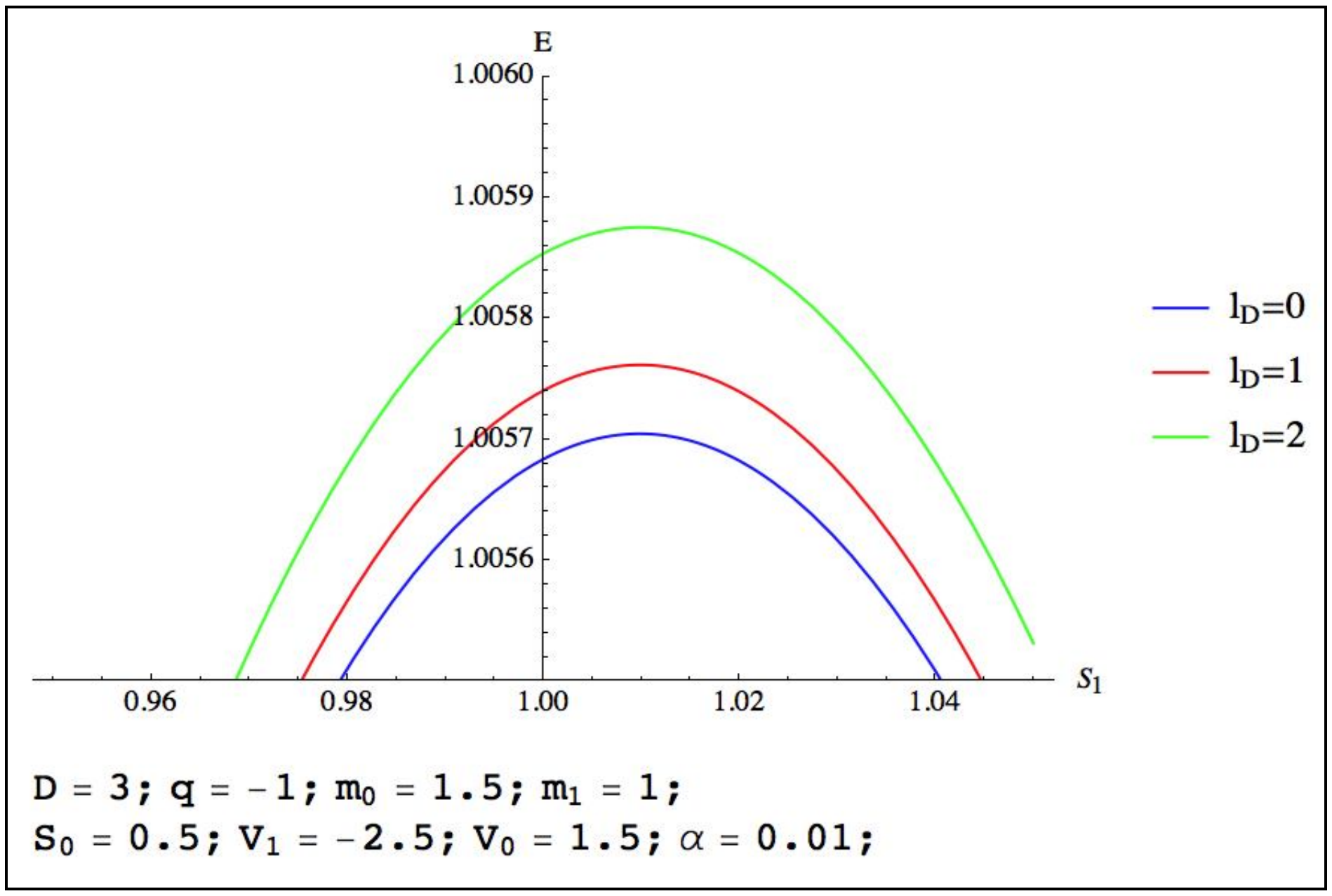

Fig.11: Plot of the energy with $S_1$ for different values of $l_D = 0{,}1$ and 2 respectively.